\documentclass{aa}
\usepackage{times} % dazugetan am 11.1.99
\usepackage{psfig}
\usepackage{epsfig}
\usepackage{epsf}
\begin{document}

   \thesaurus{(08.02.1;  % Stars, binaries
               08.16.5;  % Stars, pre-main sequence
               08.22.3;  % Stars, variable
               08.09.2 Haro 6-10;  % Stars, individual
               13.09.6)}  % Infrared, stars
   \title{The near-infrared and ice-band variability of Haro 6-10}

%   \subtitle{I. Overviewing the $\kappa$-mechanism}

   \author{Ch. Leinert
          \inst{1}
          \and
          T.L. Beck\inst{2,3}
          \and
          S. Ligori\inst{1}
          \and
          M. Simon\inst{2,3}
          \and
          J.Woitas\inst{1}
          \and
          R. R. Howell\inst{4}
        \fnmsep\thanks{Based in part on data obtained
               on the Calar Alto 3.5 m, the IRTF, UKIRT, WIRO, TIRGO,
                and the La Palma Herschel telescopes.}
          }

   \offprints{Ch. Leinert}

   \institute{Max-Planck-Institut f\"ur Astronomie, K\"onigstuhl 17,
              D-69120 Heidelberg, Germany\\
              email: leinert@mpia-hd.mpg.de
         \and
             State University of New York at Stony Brook,
                Stony Brook, NY 11974-3800, USA\\
%             email: tracy@hilo.ess.sunysb.edu
         \and
             Visiting Astronomer at the Infrared Telescope Facility, which is operated by the University of Hawaii under contract to the National Aeronautics and Space Administration.
         \and
             University of Wyoming, Laramie, WY 82071, USA
%             \thanks{The university of heaven temporarily does not
%                     accept e-mails}
             }

   \date{Received December 1, 2000; accepted ***}

   \maketitle

   \begin{abstract}

We have monitored the angularly resolved near infrared and 3.1 $\mu$m 
ice-band flux of the components of the young binary Haro 6-10 on 23 occasions
during the years 1988 to 2000.  Our observations reveal that both 
the visible star Haro 6-10 
(Haro 6-10S) and its infrared companion (Haro 6-10N) show significant 
variation in flux on time scales as short as a month. 
The substantial
flux decrease of Haro 6-10S over the last four years carries the
reddening signature of increased extinction.  However, a comparable K-band flux increase observed in the IRC is associated with a dimming in the H-band and cannot be explained by lower extinction.  Absorption in the 3.1 $\mu$m water-ice feature was always
 greater towards the IRC during our observations, indicating a larger amount of obscuring material along its line of sight.  We detect variability in the ice-band absorption towards Haro 6-10S and Haro 6-10N, significant at the 3.5$\sigma$ and 2.0$\sigma$
 levels, respectively.

   \end{abstract}

\section{Introduction} % ----------------------------------------------- 1

Haro 6-10 is a prominent member of the small class of young binaries with 
infrared companions (Koresko et al. 1997, hereafter K97).  Infrared 
companions (IRCs) to
T Tauri stars are characterized by faintness or non-detection in the visible,
very red spectral energy distributions (SEDs), and strong variability in the
infrared.  The binary system consists of Haro 6-10S, a K3-5 star (Goodrich 
1986) seen in visible light through A$_{V}$ $\approx$ 5.6 mag extinction, and 
Haro 6-10N, the IRC, 1.$''$2 to the north (Leinert \& Haas 1989a).  
Observations of the unresolved system show that its near infrared flux can 
vary significantly on timescales of months (Elias 1978, Leinert et al. 1996).

The infrared excess of Haro 6-10 is very pronounced compared to other T Tauri 
stars.  Its unresolved spectral SED is flat or rising between 1 and 
100 $\mu$m, making Haro 6-10 a class I or ``protostellar'' T Tauri star
in the system of Lada (1987).  Haro 6-10 is unique among the infrared 
companion systems because angularly resolved observations indicate that 
both components have this classification (K97).  The IRC, however, dominates 
the flux of the system at wavelengths longward of 3.5 $\mu$m and has an 
integrated luminosity twice that of Haro 6-10S (K97).  

Angularly resolved observations of the 10 $\mu$m silicate absorption feature 
indicate stronger extinction along the line of sight to the IRC (van Cleve et 
al. 1994).  Whittet et al. (1985) list the unresolved Haro 6-10 
system among the young stellar sources showing
a deep 3.1 $\mu$m water-ice absorption feature.  Angularly 
resolved observations revealed that the ice-band absorption was stronger 
towards the IRC (Leinert et al. 1996).  Apparently the lines of sight to 
Haro 6-10S and its IRC contain very different amounts of absorbing material 
even though their projected separation on the sky is only $\approx$ 170 AU.

Observations made to date have not successfully explained the cause of the 
photometric variability and extinction observed in the Haro 6-10 system.  
K97 proposed that IRCs are otherwise normal young stars experiencing strong, 
episodic extinction and variable accretion perhaps stimulated by the orbital 
motion of the binary.  We have monitored the near infrared flux and ice-band 
absorption of both components of Haro 6-10 at every opportunity.  
Our observations provide new insights to the variability of Haro 6-10 
and its IRC.

% ----------------------------------------------------Table 1

\begin{center}
\begin{table*}
\caption{\bf Journal of observations}
\label{tab:journal}
\begin{tabular}{lllllc}
\multicolumn{3}{c}{ } \\
\hline
&&&&&\\
Date & Type & Result & Telescope & Instrument & Reference\\
&&&&&\\
\hline
&&&&&\\
22 Sep. 1986 & 1D speckle & K$_{sys}$, L$'$$_{sys}$ &
        Calar Alto & own, InSb & 1 \\
11 Sep. 1987 & 1D speckle & H$_{sys}$ &
        Calar Alto & own, InSb & 1 \\
25-27 Sep. 1988 & 1D speckle & H, K, L$'$, M$_{s}$ &
        Calar Alto & own, InSb & 1 \\
11/16 Oct. 1989 & 1D speckle & K, ice, dust, L$'$ &
        Calar Alto & own, InSb & 2 \\
19-27 Sep. 1991 & 1D speckle & H, K, ice, dust, L$_{n}$, L$'$ &
        Calar Alto & own, InSb & 2 \\
28 Sep. 1991 & lunar occultation& K &
        La Palma & special, InSb & 3 \\
29 Oct. 1991 & 2D speckle & H, K, ice, dust, L$'$ &
        Calar Alto & 1-5 $\mu$m camera & 2 \\
06-10 Jan. 1993 & 1D speckle & H, K, ice, dust, L$_{n}$, L$'$, M$_{s}$ &
        Calar Alto & own, InSb & 2 \\
28 Jan. 1994 & 2D speckle & K &
        Calar Alto & Black MAGIC &   \\
15 Dec. 1994 & 2D speckle & K &
        Calar Alto & Blue MAGIC &   \\
08 Oct. 1995 & 2D speckle & K &
        Calar Alto & Blue MAGIC &   \\
26 Mar. 1996 & 1D speckle & K &
        Calar Alto & own, InSb &   \\
%%%%25 Sep. 1996 & Imaging    & J, H, K &
%%%%    TIRGO      & InSb       &  4 \\
27 Sep. 1996 & 2D speckle & K &
        Calar Alto & Blue MAGIC &   \\
30 Sep. 1996 & 1D speckle & K, ice, dust, L$_{n}$ &
        Calar Alto & own, InSb &   \\
14/15 Nov. 1997 & 1D speckle & H, K, ice, dust, L$_{n}$, L$'$, M$_{s}$ &
        Calar Alto & own, InSb &  \\
06 Mar. 1998 & 2D Speckle & K, L$'$ & IRTF & NSFCam & \\
14 Sep. 1998 & Imaging & J$_{sys}$, H, K, L$'$ & IRTF & NSFCam & \\
10 Dec. 1998 & Imaging & J$_{sys}$, H, K, CVF filters$^*$, L$'$ & IRTF &
NSFCam & \\
02 Sep. 1999 & 2D speckle & J$_{sys}$,H$_{sys}$, K$_{sys}$ &
        Calar Alto & $\Omega _{Cass}$ &  \\
18 Sep. 1999 & Imaging &  H$_{sys}$, K, L$'$, M$''$ & WIRO & IoCam 1 &
\\
08 Oct. 1999 & Imaging & H$_{sys}$, K, L$'$, M$''$ & WIRO & IoCam 1 & \\

03 Nov. 1999 & Imaging &  K, CVF filters$^*$, L$'$ & IRTF & NSFCam & \\
06 Dec. 1999 & Imaging & J$_{sys}$,  H, K, ice, dust, L$'$, M$_{s}$ &
        UKIRT & TUFTI & \\
26 Jan. 2000 & Imaging & K, L$'$ & IRTF & NSFCam & \\
07 Mar. 2000 & Imaging & K, CVF filters$^*$, L$'$ & IRTF & NSFCam & \\

&&&&&\\
\hline
\end{tabular}
\vspace*{0.3cm}\mbox{   }\\
$^1$ Leinert and Haas \cite{leinert89b}\\
$^2$ Leinert, Haas and Weitzel \cite{leinert96}\\
$^3$ Richichi, Leinert, Jameson and Zinnecker \cite{richichi94}\\
%%%$^4$ Andrea Richichi, private communication\\
$^*$ IRTF CVF filters are centered at 2.4, 2.9, 3.05 and 3.4 $\mu$m
\end{table*}
\end{center}
% ------------------------------ End Table 1

\section{Observations} % ----------------------------------------------- 2

Our data (Table 1), consisting of a heterogenous set of observations obtained
at different telescopes with different techniques, required standardisation 
in calibration. We based the calibration on the flux values for Vega given by
Tokunaga (1999), and interpolated smoothly to the wavelengths of those 
filters not covered by his Table 7.5. The IRTF data were referenced to a 
magnitude scale in which Vega has zero magnitude at all wavelengths shortward 
of 20 $\mu$m, as given by Tokunaga.  In this framework, the original flux
calibration of the Calar Alto data was on a scale in which a star of 
magnitude of 0.06 $\pm$ 0.01 would have the Vega flux listed by Tokunaga.  
Magnitudes were determined using standards from the list of Elias~et~al. 
(1982).  To have the same scale for both data sets, we adjusted the Calar 
Alto fluxes downwards by 6\%. The magnitudes are not affected by this 
recalibration. The overall consistency of the calibration of our data 
and the precision of our photometry should be better than $\pm$5~\%.

For the speckle observations (\S 3.1 and 3.2), the calibration was applied to 
the total system flux.  The fluxes of the components were obtained 
subsequently using their flux ratio derived from analysis of the speckle 
observations. For flux ratios smaller than 0.10, an additional 
uncertainty up to 10\%  enters the flux determination of the faint component,
 because the average 
error for the determination of the flux ratio in H or K is $\pm$ 0.007. 
Calibration of the imaging observations (in particular \S 3.3) was applied 
directly to the
components of Haro 6-10.

% ----------------------------------------------------- Table 2
\begin{center}
\begin{table*}
\caption{Individual magnitudes and colours for the components.  
        A  subscript S is for the visible star Haro 6-10S, and N is for the 
        IRC, \mbox{Haro 6-10N.}}
\label{tab:magnitudes}
\begin{tabular}{rlllllllllllll}
\multicolumn{13}{c}{ } \\
\hline
&&&&&&&&&&&&\\
No. &Date & H$_{N}$ & H$_{S}$ & K$_{N}$ & K$_{S}$ & H-K$_{N}$ & H-K$_{S}$ &
L$'_{N}$ & L$'_{S}$ & K-L$'_{N}$ & K-L$'_{S}$ & M$_{N}$ & M$_{S}$ \\
&&&&&&&&&&&&\\
\hline
&&&&&&&&&&&&\\
1& 25-27 Sep. 1988 & 11.88 & 8.99 & 9.92 & 7.70 & 1.96 & 1.29 & 6.49 & 6.18 & 
3.43 & 1.52 & 5.02 & 5.26 \\
2& 11/16 Oct. 1989 & -- & -- & 9.76 & 7.30 & -- & -- & 6.64 & 5.73 & 3.12 & 
1.63 &-- & -- \\
3& 19-27 Sep. 1991 & 11.73 & 8.54 & 10.42 & 7.42 & 1.31 & 1.12 & 6.87 & 5.98 & 
3.55 & 1.46 & -- & -- \\
4& 28 Sep. 1991 & -- & -- & 9.60 & 7.10 & -- & -- & -- & -- & 
-- & -- & -- & -- \\
5& 29 Oct. 1991 & 11.82 & 8.50 & 10.34 & 7.39 & 1.48 & 1.11 & 7.13 & 5.67 & 
3.20 & 1.72 & -- & -- \\
6& 06-10 Jan. 1993 & 12.55 & 8.63 & 9.84 & 7.34 & 2.71 & 1.29 & 6.35 & 5.85 & 
3.49 & 1.49 & 4.82 & 4.95 \\
7& 28 Jan. 1994 & -- & -- & 11.00 & 7.30 & -- & -- & -- & -- & -- & -- & -- & 
--\\
8& 15 Dec. 1994 & -- & -- &  9.54 & 7.21 & -- & -- & -- & -- & -- & -- & -- & 
--\\
9& 08 Oct. 1995 & -- & -- & 10.24 & 7.09 & -- & -- & -- & -- & -- & -- & -- & 
--\\
10& 26 Mar. 1996 & -- & -- &  9.82 & 7.18 & -- & -- & -- & -- & -- & -- & -- &
 --\\
11& 27 Sep. 1996 & -- & -- & 10.32 & 7.54 & -- & -- & -- & -- & -- & -- & -- &
 --\\
12& 30 Sep. 1996 & -- & -- & 9.91 & 7.39 & -- & -- & -- & -- & -- & -- & -- & 
--\\
13& 14/15 Nov. 1997 & 12.62 & 9.82 & 10.19 & 8.05 & 2.43 & 1.75 & 6.40 & 6.07 
& 3.79 & 1.98 & 4.62 & 5.24 \\
14& 06 Mar. 1998 & -- & -- & 9.97 & 8.89 & -- & -- & 6.01 & 6.26 & 3.96 & 
2.63 &-- & -- \\
15& 14 Sep. 1998 & 12.81 & 10.69 & 10.14 & 9.08 & 2.67 & 1.61 & 5.79 & 6.32 &
4.35 & 2.76 & -- & -- \\
16& 10 Dec. 1998 & 12.39 & 10.70 & 9.99 & 9.12 & 2.40 & 1.58 & 5.99 & 6.39 
& 4.00 & 2.73 & -- & -- \\
17& 02 Sep. 1999 & 12.37 & 10.24 & 9.24 & 9.52 & 3.13 & 1.22 & -- & -- & -- &
-- & -- & -- \\
18& 18 Sep. 1999 & -- & -- & 9.59 & 9.43 & -- & -- & 5.46 & 6.36 & 4.13 & 3.07
& -- & -- \\
19& 08 Oct. 1999 & -- & -- & 9.31 & 9.31 & -- & -- & 5.41 & 6.16 & 3.90 & 
3.15 & -- & -- \\
20& 03 Nov. 1999 & -- & -- & 8.80 & 9.25 & -- & -- & 4.78 & 6.17 & 4.02 & 
3.74 &-- & -- \\
21& 06 Dec. 1999 & 13.52 & 10.63 & 8.99 & 8.88 & 4.61 & 1.75 & 5.23 & 6.40 & 
3.58 & 2.48 & 4.07 & 5.49 \\
22& 26 Jan. 2000 & -- & -- & 8.89 & 8.79 & -- & -- & 5.15 & 6.19 & 3.74 & 
2.60 &-- & -- \\
23& 07 Mar. 2000 & -- & -- & 8.66 & 8.61 & -- & -- & 4.91 & 6.13 & 3.75 & 
2.48 &-- & -- \\

&&&&&&&&&&&&&\\
\hline
\end{tabular}
\end{table*}
\end{center}
% --------------------------------End Table 2

\begin{figure*} % ------------------------------------------ figure 1
\vbox{\leftline{\psfig{figure=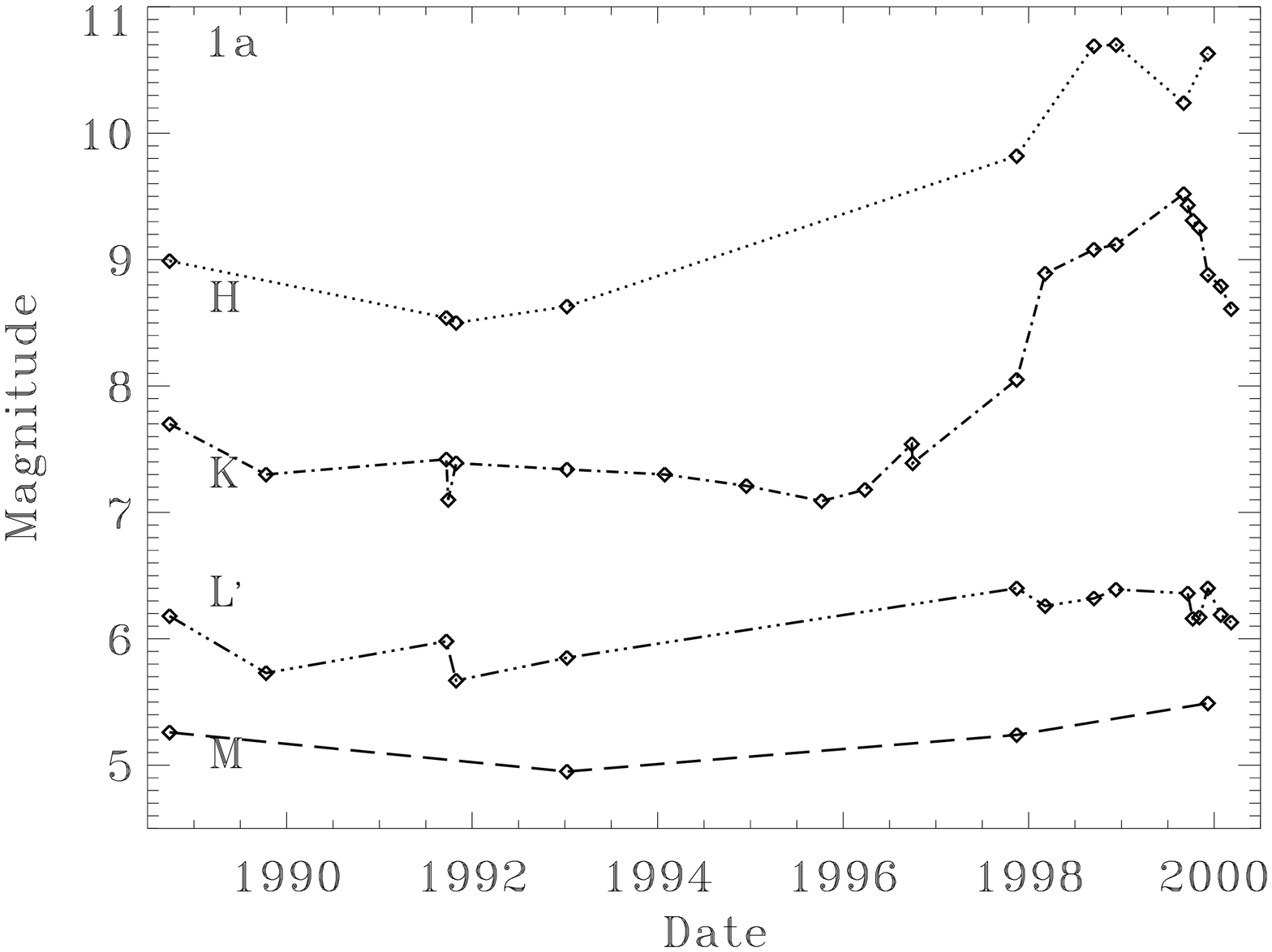,width=65mm,angle=90}}}
\vspace*{-6.5cm}
\hfill\parbox[b]{8.8cm}{\rightline{
\psfig{figure=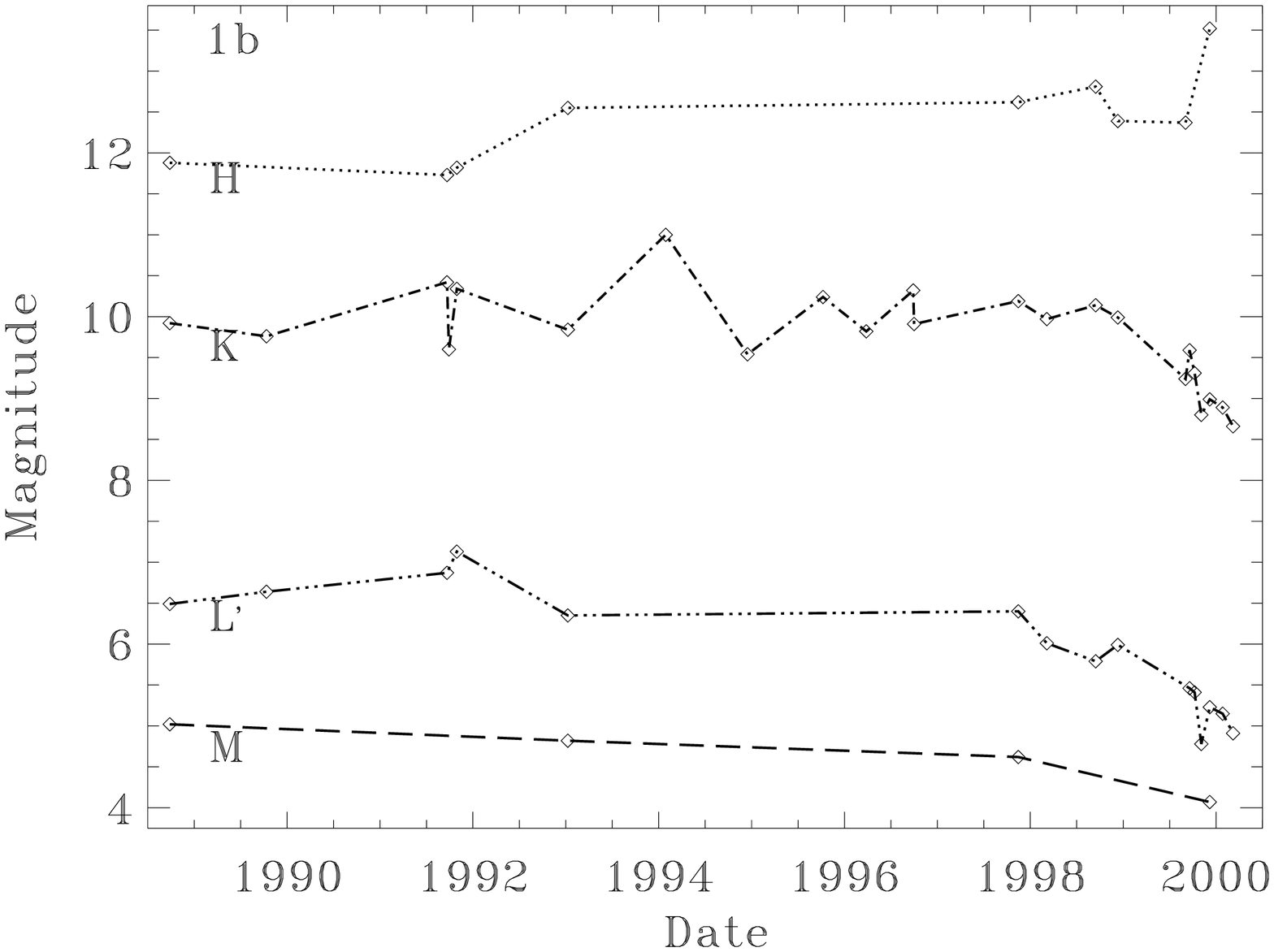,width=65mm,angle=90}}}
\caption{Variation of the near-infrared magnitude of 
Haro 6-10S (1a) and Haro 6-10 IRC (1b) from 1988 to 2000.}
\label{fig:mag_col_components}
\end{figure*}  % ------------------------ end figure 1

\section{Results} % --------------------------------------------------- 3

\subsection{Photometric Variability}% -------------------------------3.1

Table 2 lists the near IR magnitudes and colors of Haro 6-10N and S.
The IRC has been brighter than Haro 6-10S at L' and M
throughout our observations of the system.  At shorter wavelengths,
the IRC was significantly fainter than Haro 6-10S at the start of our 
observations and slowly became fainter with time (Figures 1a and b).
In 1999, however, the IRC brightened, while Haro 6-10S was in a phase
of decreasing brightness, so that in November, 1999 it was
about 50\% brighter than Haro 6-10S at K.  This is the first time
in our experience that an IRC has been brighter at K than the component
seen in visible light.

Figure 2 plots the 15 available measurements of the K-L$'$ color 
vs. K band magnitude of Haro 6-10S.  The figure shows that when 
Haro 6-10S is fainter at K, it is also redder.  The dependence of its
H-K color on K magnitude shows a similar correlation.  This has a 
natural 
explanation if changes in absorbing material along the line of 
sight are primarily responsible for the variation in its near infrared flux.
The temporal evolution of K-L$'$ {\it vs} K was such that 
the observed values lay
in the lower right part of the diagram before 1997, moved upwards to the 
left until November 1999, and again moved  downwards toward the right starting 
in December, 1999. This coordinated behavior reinforces the  suggestion that a 
one-parameter effect, probably variable extinction, is responsible for the 
observed changes in flux.  

Fig. 2 shows the best linear fit, in the least squares sense, to the 
observed correlation; its slope is 0.81$\pm$0.23.   The slope  for 
extinction following the wavelength dependence measured in the interstellar 
medium,  A$_{K}$/A$_{V}$=0.112 and A$_{L}$/A$_{V}$=0.058 (Rieke \& 
Lebofsky 1985), is 0.51.   Overplotted on Figure 2 is a line of this slope.  The difference in slopes could be 
attributable to a different extinction law for material around Haro~6-10S, but 
further multi-wavelength observations are necessary to determine if this 
difference is statistically significant.

Fig. 3a shows K-L$'$ vs K mag for the IRC with the data identified as in 
Fig. 2.  The figure is divided into into quadrants by lines at K-L$'$=3.55 
and K=9.7 mag.  All of the measurements before 1993 lie in the lower left
quadrant, all of the 1997/1998 measurements in the upper left quadrant, 
and the 1999/2000  measurements in the upper right corner, with the points 
moving right.  The IRC became redder while staying faint between 1993 and 
1997, and grew brighter at the approximately same colour from 1998 on. The 
net effect is one of the IRC becoming redder and brighter, which is the 
opposite from what would be expected to result from variable extinction.  
This behavior is 
even more pronounced in a plot of the H-K color vs K (Fig. 3b). 
Plotting H-K vs H (Fig. 3c) shows a more normal behaviour in that the 
reddening was accompanied by
an overall dimming at H.  But here the slope of the observed H-K vs H 
relation,
$\approx 1.2$, is much too steep to be explained by extinction alone.

\begin{figure} % -----------------------------------------------figure 2
\rightline{\psfig{figure=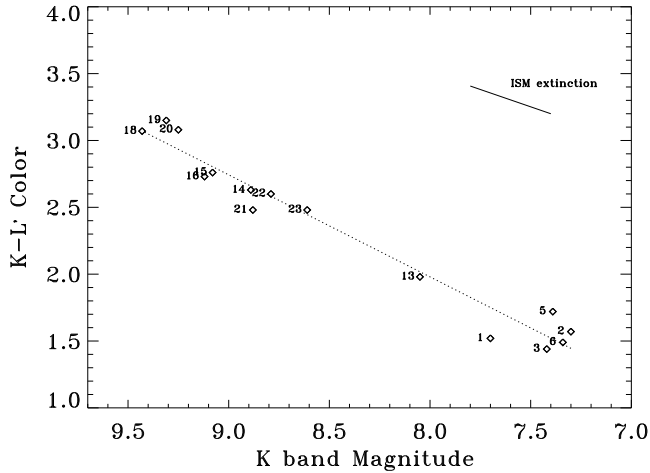,width=7.0cm,angle=90}}
\caption{K-L$'$ color vs. K-band magnitude for Haro 6-10S.  
Overplotted on the figure is a linear fit to the data (dotted line) 
and the expected slope if changes in near-infrared flux are caused by 
variable extinction (solid line).
To help identify the observation times, we identified the 
data points by the numbering introduced in the first column of Table 2.  
}
\label{fig:color_vs_mag_south}
\end{figure} % ----------------------end figure 2

The Haro 6-10 binary has now been observed in the IR for more than a quarter 
century. Figure 4 shows the light curve of the total system in the H, K, L', 
and
M photometric bands.  The measurements before 1988 are of the total,
unresolved system; those before 1978 are from Elias (1978), those for
1981 from Cohen and Schwartz (1983), and for 1984 from Myers et al. (1987).
For the later data, the values are the sums of the component entries
in Table 2.  C. Koresko and A. Richichi (priv. comm.) provided the values 
for 1992 and additional data for September 1996.
The decrease of total K-band flux reported  by Elias (1978) was
accompanied by reddening of the (H-K) color.  This suggests that Haro 6-10S 
dominated the NIR flux of the system then, as it did during most of our 
observations, and that increased obscuration was responsible for its 
dimming.  If we assume that indeed Haro 6-10S dominated the 
light of the system 
at H and K during the entire time before 1988 when 
only integrated measurements were available, then Haro 6-10S  decreased in 
K-band flux by about a factor of 10 over the past 25 years , with major events,
decreases by a factor of about three, during the first two and the
last two years of this period. In contrast, the L' and M bands dominated 
by the IRC show a gradual increase over the period of observations and a
particularly strong increase during the past two years. This different 
behaviour of the light  curve at short and long wavelengths also suggests 
that different physical phenomena are responsible for the variability of
Haro 6-10S and the IRC.

% ----------------------------------------------------------------- Figure 3
\begin{figure*}
\vbox{\psfig{figure=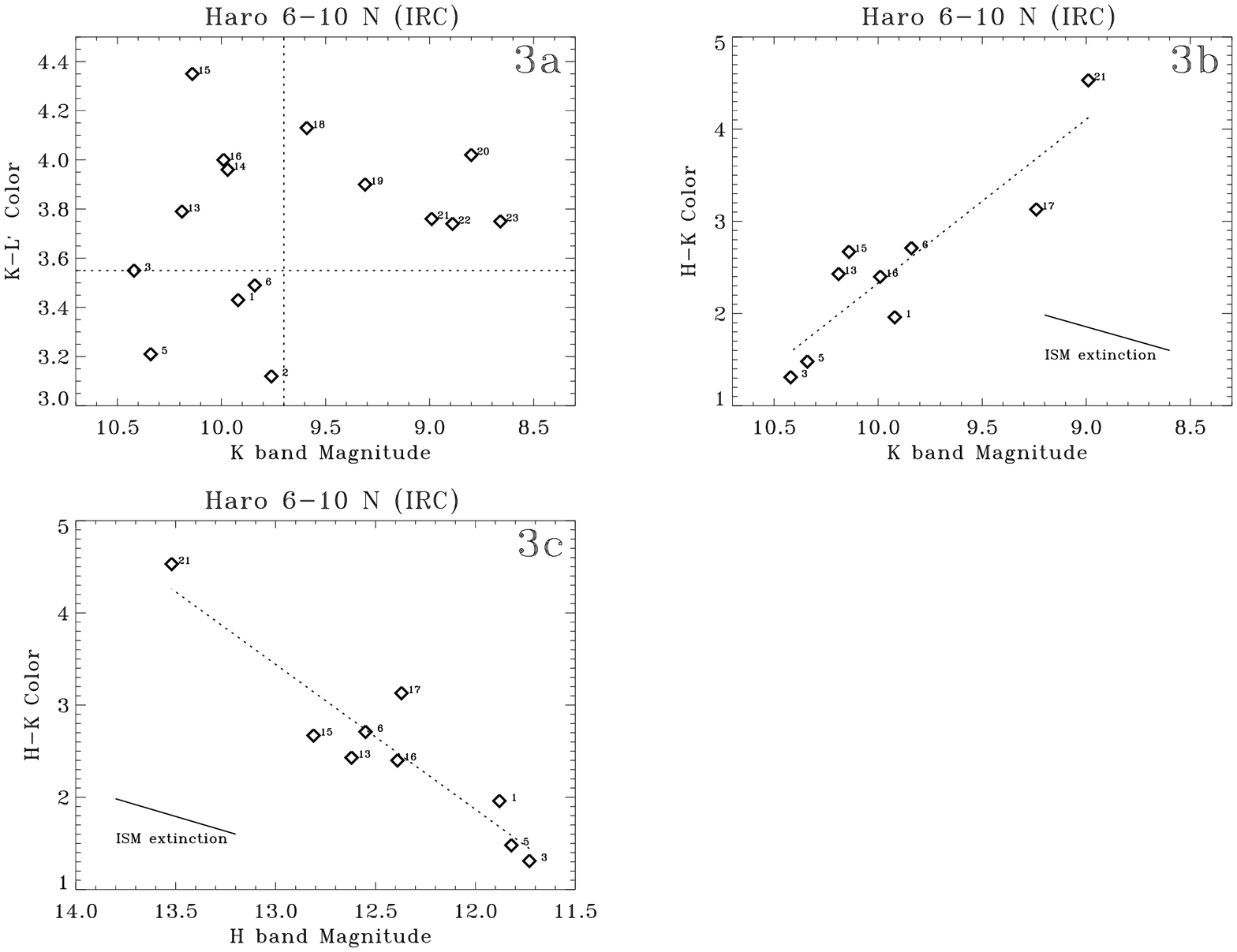,width=13.5cm,angle=90}\vspace{-5.6cm}}
\hspace*{10.5cm}\parbox[b]{7.5cm}{
\caption{3a presents the K-L$'$ color vs. K-band magnitude for the 
Haro 6-10 IRC.  3b is the H-K color vs. K-band magnitude plot and 3c is 
the H-K color vs. H-band magnitude.  Overplotted on 3b and 3c is a 
linear fit to the data (dotted line) and the expected slope if near 
infrared flux variation is caused by variable extinction (solid line).  
In the three plots, the data points are numbered by observation date as 
they appear in Table 2. We note that in 3a the lower left
quadrant contains data from before 1993, the upper left quadrant
data from 1997/98, and the upper right quadrant data from 1999/2000.}
        \label{fig:colour_magnitude_north}}
\vspace*{2cm}
\end{figure*}
% ---------------------------- End Figure 3

\begin{figure}[b!] % -----------------------------------------------figure 4
\leftline{\psfig{figure=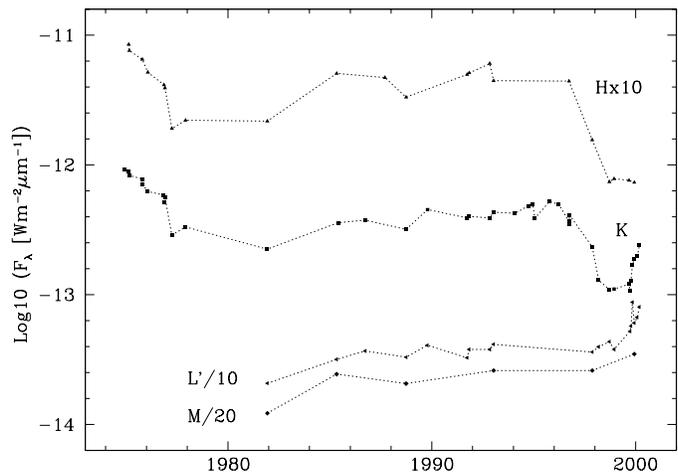,width=7.0cm,angle=-90}}
\caption{Variation of near-infrared
        flux of Haro 6-10 from 1974 to 2000. Measurements before 1978
are from Elias (\cite{elias78}), for 1981 from Cohen and 
Schwartz (\cite{cohen83}), for 1984 from Myers~et~al. (\cite{myers87}),
        for November 1992 from C. Koresko (private communication).}
\label{fig:Flux_variation}
\end{figure} % ----------------------end figure 4

\subsection{Ice-band Variability}% -------------------------------3.2

Figures 5a and 5b show measurements of spectral energy distributions 
of Haro 6-10S and the IRC centered on the 3.1 $\mu$m water-ice feature.    
  We estimated the ice-band optical depth, 
$\tau_{ice}$, by measuring the absorption with respect to a continuum that 
was drawn as a straight-line fit between the fluxes at K and L$'$  (or 
3.5 $\mu$m if no L$'$ measurement was available).  These values, and their 
uncertainties, are indicated in Figs. 5a and 5b.  The uncertainties
in these estimates are the result of the variable shape of the component
SEDs, incomplete spectral sampling of the SEDs, and imprecision of the
photometry.  We find that the optical depths to Haro 6-10S and its IRC are 
variable at the 3.5$\sigma$ and 2$\sigma$ levels, respectively.  
%%Unfortunately, the uncertainties in $\tau_{ice}$ are sufficiently large that 
%%%it is not meaningful to investigate correlations with the variations in
%%%color or flux of the components. 

The large uncertainties in the values of $\tau_{ice}$ make it difficult to
establish correlations. However, for Haro 6-10S the decrease in K brightness
after 1997 by about 1.5 magnitudes was accompanied by an increase of
$\tau_{ice}$ by about 0.3. If the brightness decrease is due to
increased extinction as we have argued in \S 3.1, then this increase 
is much less than the change in optical depth of the ice band by 1.2
expected from Whittet et al.'s (1988)
relation between A$_V$ and $\tau_{ice}$ for the interstellar medium (ISM).
This suggests again that the material around Haro 6-10 may have properties 
different from the general ISM. For the IRC no clear conclusions can be drawn, 
but high values of $\tau_{ice}$ did not occur in our measurements when
the IRC was brighter than K = 9 mag. 

It is interesting to consider the average $\tau_{ice}$ toward 
Haro 6-10S and the IRC, which have values of $0.34\pm0.06$ and $0.68\pm0.07$, respectively.
Using the $\tau_{ice}$ to A$_{V}$ relation from Whittet et al. (1988), 
$\tau_{ice}$=m(A$_{V}$-A$_{V}$(0)) where A$_{V}$(0)=3.3$\pm$0.1 and 
m=0.093$\pm$0.003, we derive average visual extinctions  $7.0\pm0.7$ 
mag for Haro 6-10S, and $10.6\pm0.8$ for the IRC.  Our result is consistent with previous angularly resolved 
observations of the silicate and water-ice absorption features (van Cleve et al. 1994, Leinert et al. 1996).

\subsection{Morphology in the Near Infrared}% --------------------------3.3

In December, 1999 we imaged Haro 6-10 at UKIRT using the near-IR camera 
TUFTI. It is equipped with a $256\times256$ pixel$^2$ InSb detector.
The optics provide a plate scale of 0.081 arcsec/pixel to take advantage
of the wavefront sensing and active optics capabilities of the UKIRT.  

\begin{figure*} % ------------------------------------------ figure 5
\vbox{\leftline{\psfig{figure=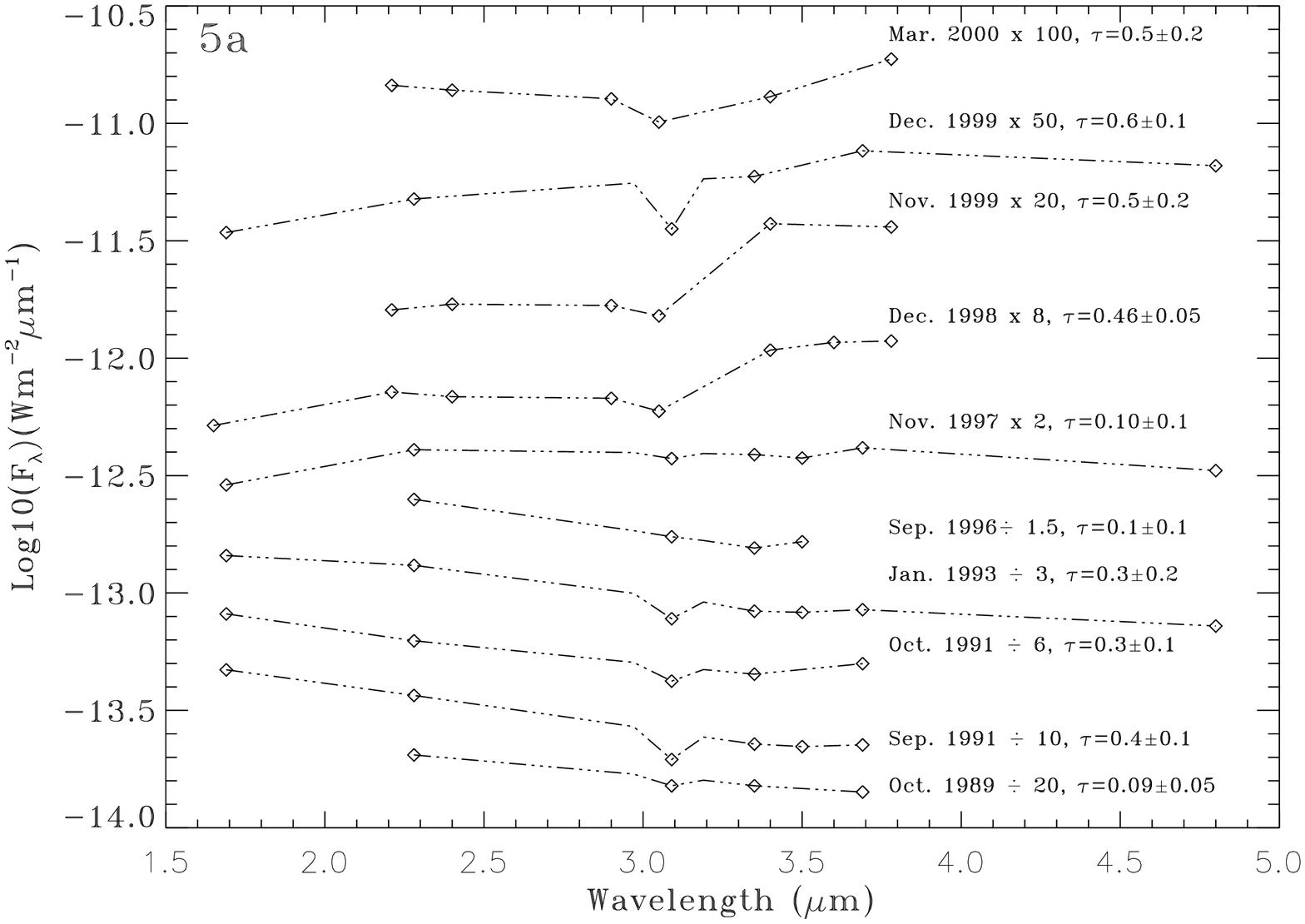,width=65mm,angle=90}}}
\vspace*{-6.5cm}
\hfill\parbox[b]{8.8cm}{\rightline{
\psfig{figure=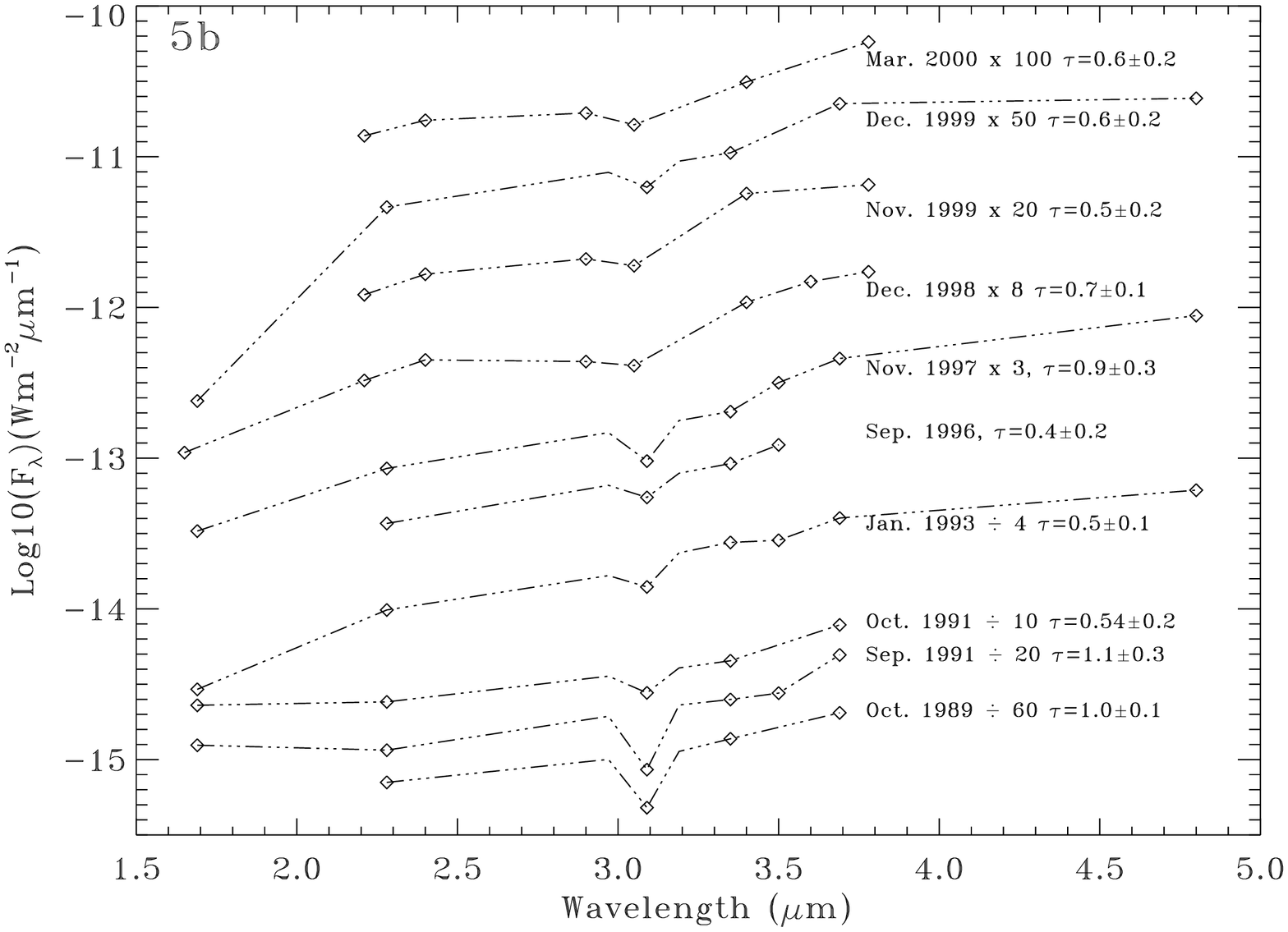,width=65mm,angle=90}}}
\caption{Variation in ice-band absorption of Haro 6-10S (5a) and its IRC (5b).}
\label{fig:ice_flux_components}
\end{figure*}  % ------------------------ end figure 5

Fig. 6 shows the images obtained in the J, H, K, ``ice'' ($3.10\mu$m),
``dust'' ($3.28\mu$m), L$'$, and M filters.  Haro 6-10S and the IRC are resolved
in all the images except J in which the IRC was not detected. At the time
of these observations the components were essentially equal in brightness
at K and the IRC brighter at longer wavelengths (see also Table 2). 
The sensitivity of the K frames is insufficient to show the faint north-south 
extensions on the IRC found by Koresko~et~al. (1999). 
The short wavelength  images show the jetlike structure toward the south 
discovered by Movsessian and Magakian (1999) in the light of the S[II] 
6716+6731 line emission and imaged in K by Koresko~et~al. (Koresko 1999).
We emphasize this structure in J and H images by suitable choice of
the scaling. Obviously, scattered light contributes strongly to the 
brightness of this feature and its classification as collimated supersonic 
outflow therefore requires further study. Another interesting 
feature, best visible at H, is an arc of emission 6$^{\prime\prime}\!$.2
E of the binary system, extending roughly from position angle 90$^{\circ}$
to 120$^{\circ}$ . The arc, marginally visible on the image of
Koresko~et~al. (1999), is apparently connected to the nebulosity 
closer to the system both at its northern and southern end. Spectral line 
imaging is necessary to clarify if this is another bow shock, similar to 
HH~184 F and G shown in the paper of Devine~et~al. (1999).  If it is a bow
shock, then it defines a new outflow system, at PA $\approx$ 110$^\circ$ and adds 
one more complexity to the immediate environment of Haro~6-10.  At longer 
wavelengths (L$'$ and M filters), emission from the infrared companion becomes 
dominant with respect to the southern component and the 
jetlike feature is no longer detected. Despite the substantial brightness
changes of the components of Haro 6-10 there were no significant changes
in morphology.

\section{Discussion} %------------------------------------------4 

Our results show convincingly,
by the near quantitative agreement with the reddening predictions
for such a model, that the NIR flux variability of the star seen 
in visible light, Haro 6-10S, is caused in large 
part by variations in extinction. 
The average extinction measured towards Haro 6-10S using the ice-band is 
in good agreement 
with K97's estimate, A$_{V}$ = 5.6 mag,
if we allow for the fact that our observations included a period of
higher extinction.  Our data do not 
provide information directly about where the obscuring material lies and 
why its column density varies. The variability 
is probably attributable to inhomogeneities in the obscuring material and their 
motion across our line of sight.  Since this line of sight contains 
sufficient material to produce A$_{V}$ = 5 to 10 mag, but yet is sufficiently
transparent to permit detection of the star, it may prove useful for
a comparison of the composition of material near a young star with that
in the interstellar medium.  For example, it should be possible to compare
the ratios of the water ice and silicates in the two environments.

% ----------------------------------------------------------------- Figure 6
\begin{figure*}
\vbox{\psfig{figure=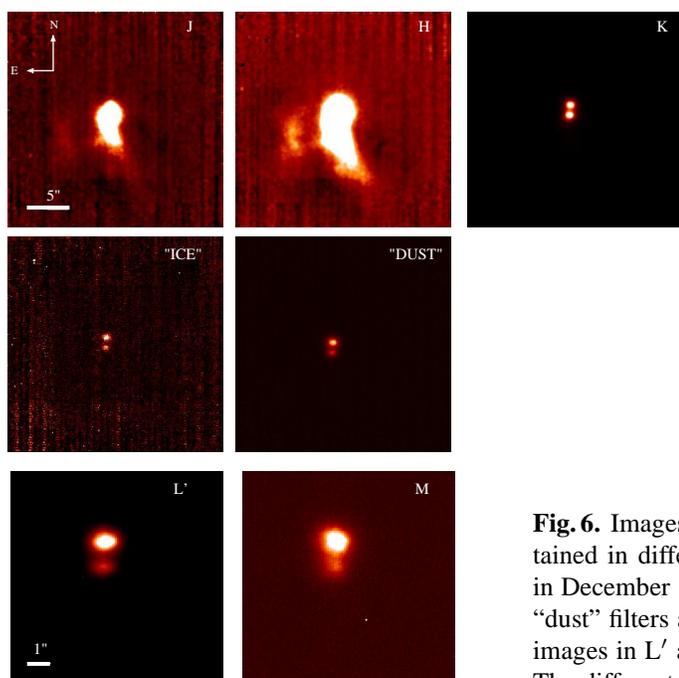,width=9.0cm,angle=0}\vspace{-2.6cm}}
\hspace*{7cm}\parbox[b]{7.5cm}{
\caption{Images of Haro 6-10 and its
        companion obtained in different near-infrared bands from
        UKIRT in December 1999.
        The images in J, H, K, ``ice'', 
        and ``dust'' filters are a mosaic of 5 dithered images. The 
        images in L$^\prime$ and M filters represents a single frame. 
        The different scales are shown in the J and in the 
        L$^\prime$ images.}
        \label{fig:UKIRT_images}}
\vspace*{1cm}
\end{figure*}
% ---------------------------- End Figure 6

Changes in accretion rate do not appear to be a main cause for the
brightness variations of Haro 6-10S. From the example given by 
Calvet~et~al. (\cite{calvet97}) we find the near-infrared colours H-K and 
K-L$'$ almost independent of accretion rate over the range of 
10$^{-8}$M$_{\odot}$/yr to 10$^{-6}$M$_{\odot}$/yr. Emission from an
infalling envelope, as probably present in highly veiled sources,
tends to redden the energy distribution with increasing accretion
luminosity, which is not what we observe. Quantitative predictions
of these intricate effects would need specific modelling of the source
and is beyond the scope of our paper.

The flux variability of the IRC is much more complex than that of Haro 6-10S.
The extinction implied by the ice-band absorption is much smaller than
K97's estimate of A$_{V}$ $\approx$ 49 mag derived on the assumption that 
the IRC is
a strongly obscured young star.  A certain discrepancy  in the
derived A$_V$ is not too surprising
because the relatively large value of $\tau_{ice}$ suggests that the line
of sight could be optically thick in the ice-band.  Radiative transfer 
effects and scattering
could then produce an underestimate of the actual extinction to a 
star powering the IRC.  But more significantly,
the color variations of the IRC (e.g. Fig 3) indicate that  processes other 
than, or in addition to, extinction must be involved in its variability.
Interpretation of reddening by extinction therefore may not be
adequate for Haro 6-10N. 
Herbst et al.'s (1995) detection of H$_2$ toward the IRC but not
Haro 6-10S indicates the presence of shocked material, which points to 
accretion as an additional process playing a role in the IRC variations.
Muzzerole et al.
(1998) have demonstrated the usefulness of Br$\gamma$ emission to measure
the accretion luminosity.  A time series of angularly resolved measurements 
of H$_2$, Br$\gamma$ and Br$\alpha$ emission of the IRC is likely to 
provide a useful 
diagnostic of its activity and provide further insights to its nature,
in particular when coupled to monitoring of extinction indicators like
reddening and dust absorption fetures.

\newpage
\section{Summary}%------------------------------------------5

Our angularly resolved NIR observations of the Haro 6-10 binary show
that:

1.  Both components vary significantly in near infrared flux on timescales 
of a few months.

2.  The near infrared variability of Haro 6-10S, the visible light star, 
is caused mostly by changes in extinction.  On the other hand, the near 
infrared variability of the IRC cannot be explained solely by variable 
extinction.

3.  Absorption by water-ice is present in both components and is consistently 
stronger toward the IRC.  There is evidence that the ice-band absorption is 
variable in time toward both components.
 
\begin{acknowledgements}

We acknowledge with pleasure the hospitality provided by Hans Zinnecker
and the organizers of IAU Symposium 200 where this paper was conceived.  
We thank C. Koresko and A. Richichi
for allowing us to use their 1992 and September 1996 data, and D. 
Griep and P. Fukumura-Sawada for obtaining the service observations at the 
IRTF in March 2000. The work of TB and MS was supported in part by NSF Grant 
AST98-19694.

\end{acknowledgements}

\end{document}